\documentclass[12pt,a4paper]{article}    

\usepackage{graphicx,amsfonts}

\newcommand{\be}{\begin{equation}}
\newcommand{\ee}{\end{equation}}        	
\newcommand{\bea}{\begin{eqnarray}}
\newcommand{\eea}{\end{eqnarray}} 
\newcommand{\ba}{\begin{align}}
\newcommand{\ea}{\end{align}}  			

\newcommand{\medspace}{\:}

\newcommand{\hugenegspace}{\!\!\!\!\!\!\!\!\!\!\!\!\!\!\!\!\!\!\!}

\newcommand{\s}{{}^* \negthinspace}

\newcommand{\cd}[2]{{#1}{}_{;#2}}

\newcommand{\cds}[2]{{#1}{}_{\s ;#2}}

\newcommand{\M}{\ensuremath{\mathcal{M}}}
\newcommand{\U}{\ensuremath{\mathcal{U}}}

\newcommand{\R}{\ensuremath{\mathbb{R}}}

\newcommand{\plus}{\mathop{\!+\!}}
\newcommand{\minus}{\mathop{\!-\!}}

\newcommand{\smallfrac}[2]{\mbox{\small $\frac{#1}{#2}$}}

\renewcommand{\vec}{\textbf}

\newcommand{\text}[1]{\mbox{#1}}

\newcommand{\qed}{\rule{3mm}{3mm}}

\newcommand{\implies}{\Rightarrow}

\newcommand{\nrightarrow}{\rightarrow \!\!\!\!\!\! / \medspace\ }

\newenvironment{ind}{}{}
\newenvironment{proof}[1]{\noindent\textbf{#1:}\\}{\hfill\qed}

\newtheorem{thm}{Theorem}[section]
\newtheorem{defn}{Definition}[section]
\newtheorem{conj}{Conjecture}[section]

\begin{document}

\title{\textbf{Isotropic singularities in shear-free perfect fluid cosmologies}}
\author{Geoffery Ericksson
	\thanks{Department of Theoretical Physics, RSPhysSE, IAS, 
	Australian National University, Canberra, ACT 0200, Australia
	\quad email: Geoffery.Ericksson@anu.edu.au} 
	\and Susan M. Scott
	\thanks{Department of Physics and Theoretical Physics, 
	Faculty of Science, 
	Australian National University, Canberra, ACT 0200, Australia
	\quad email: Susan.Scott@anu.edu.au}}

\maketitle			

\begin{abstract}
We investigate barotropic perfect fluid cosmologies which admit an isotropic 
singularity.  From the General Vorticity Result of Scott, it is known that 
these cosmologies must be irrotational.  In this paper we prove, using two 
different methods, that if we make the additional assumption that the perfect 
fluid is shear-free, then the fluid flow must be geodesic.  This then implies 
that the only shear-free, barotropic, perfect fluid cosmologies which admit an 
isotropic singularity are the FRW models.
\end{abstract}


\section{Introduction}

In 1985 Goode and Wainwright \cite{GW85} introduced the concept of the 
isotropic singularity (IS) (see Appendix \ref{ap:defns} for the definition) to the field of mathematical cosmology in order to 
clarify what is meant by a 
``Friedmann-like'' singularity.  A question which naturally arises is the
following: ``precisely what cosmologies actually admit an isotropic 
singularity?''  In order to make the
 problem more tractable, the specialisation to perfect fluids has been studied 
by various authors with some success.  Goode \cite{Goode87} has shown that 
perfect fluid cosmologies satisfying the dominant energy condition, with a 
$\gamma$-law equation of state, which admit an 
isotropic singularity, must be irrotational.  Scott \cite{SSGVR} has extended 
this result to include all barotropic perfect fluids satisfying the dominant 
energy condition --- this is known as the 
General Vorticity Result (GVR).  In this paper we will enlarge the current 
knowledge of perfect fluid cosmologies which admit an isotropic singularity to
 include the following theorem.

\begin{thm}[Zero Acceleration Result]
\label{thm:zar}
If a space-time $(\M,g)$:
\begin{itemize}
\item is a $C^3$ solution of the Einstein field equations (EFE), 
\item has barotropic perfect fluid source, 
\end{itemize}
and the unit timelike fluid congruence \vec{u} is:
\begin{itemize}
\item shear-free,
\item regular at an isotropic singularity (with $-1<\lambda<1$), 
\end{itemize}
then the fluid flow is necessarily geodesic.
\end{thm}

We note that the technical condition $-1<\lambda<1$ implies that the barotropic perfect fluid cosmology satisfies the dominant energy condition.  The dominant energy condition can be used instead of the technical condition $-1<\lambda<1$ in the statement of Theorem \ref{thm:zar}.  In the interests of brevity, it will henceforth be assumed that when we say ``barotropic perfect fluid cosmology'' we actually mean ``barotropic perfect fluid cosmology satisfying the dominant energy condition''.

The General Vorticity Result enables us to say that barotropic perfect fluid 
cosmologies with non-zero vorticity do not admit an IS.  Theorem \ref{thm:zar}, 
which we will refer to as the Zero Acceleration Result (ZAR), implies that a 
shear-free, barotropic perfect fluid cosmology with non-geodesic fluid flow, 
also does not admit an IS.  Since the Friedmann-Robertson-Walker (FRW) 
cosmologies are characterised by their globally vanishing vorticity, shear, 
and acceleration, the GVR and ZAR can be combined to produce the fact that 
\emph{shear-free, barotropic, perfect fluid cosmologies which are not FRW models, do 
not admit an IS.}

The following conventions are used throughout this paper.
\begin{itemize}
\itemsep=0pt
\parsep=0pt
\item Latin letters denote 0,1,2,3.  Greek letters denote 1,2,3. 
\item $\s{}\medspace$ denotes that the entity to which it is attached exists
	 in the unphysical space-time $(\s\M,\s g)$, not the physical 
	space-time $(\M,g)$.
\item $;$ denotes the covariant derivative w.r.t.\ the physical metric $g$ .
\item $\s ;$ denotes the covariant derivative w.r.t.\ the unphysical metric 
	$\s g$ .
\item A dot above an entity in the physical space-time means that we take the 
	covariant derivative of the entity w.r.t.\ the physical metric $g$ in
	 the direction of the fluid 
	velocity field, $\vec{u}$,  i.e., $\dot{A}=\cd{A}{a}u^a$.  
	Similarly, a dot above 
	an entity in the unphysical space-time means that we take the 
	covariant derivative of the entity w.r.t.\ the unphysical 
	metric $\s g$ in the direction of the unphysical 
	fluid velocity field, $\s \vec{u}$, i.e., 
	$\s\dot{A}=\cds{\s A}{a}\s u^a$.
\item $A=O(B)$ means $-K\leq \frac{A(x)}{B(x)}\leq K$ as $x\rightarrow x_0$, 
	for some $K\in\R^+$.
\item $A=o(B)$ means $\frac{A(x)}{B(x)}\rightarrow 0$ as $x\rightarrow x_0$.
\item Two functions, $A$ and $B$, are said to be 
	\emph{asymptotically equivalent}
	 as $x\rightarrow x_0$, written $A(x)\approx B(x)$,  if 
	$A(x)=B(x)\{1+o(1) \}$ as $x\rightarrow x_0$.
\end{itemize}


\section{Proof of the Zero Acceleration Result by elimination}
\label{sec:proof_of_zar_elimination}

We proceed to prove the Zero Acceleration Result (ZAR) using two different 
methods.  In this section we present a proof by elimination, based on a theorem
 due to Collins and Wainwright \cite{CW83}.  The importance of this method of 
proof, given that we provide a general proof of the ZAR in Section 
\ref{sec:general_proof_of_zar}, lies in the demonstration of the numerous techniques 
that we now have at our disposal for deciding whether or not a specific given 
space-time actually has an IS.  We commence by quoting the Collins and 
Wainwright theorem.

\begin{thm}
\label{thm:CW83}
Any barotropic, irrotational, shear-free, perfect fluid solution of the EFE with 
non-zero expansion and $\mu +p \neq 0$ is either  
\begin{enumerate}
\item  a FRW model, or
\item  a spherically symmetric Wyman model, or
\item  a plane symmetric Collins-Wainwright model.
\end{enumerate}
\end{thm}

\begin{proof}{Proof of the ZAR by elimination}	

Scott \cite{SSGVR} has shown (GVR) that a space-time which satisfies the 
conditions of Theorem \ref{thm:zar} must be irrotational.  It is also 
shown in \cite{SSGVR} that if a perfect fluid space-time has an isotropic 
singularity at which the unit timelike fluid congruence is regular, then 
$\mu+p\neq 0$ as the singularity is approached.  
(More precisely, if a perfect fluid 
space-time has an IS at which the unit timelike fluid congruence is regular,
 then there exists an open neighbourhood $\U$ of the 
spacelike hypersurface $T=0$ in $\s\M$ such that $\mu+p\neq 0$ anywhere in 
$\U\cap\M$.)  Finally, we note that Goode and Wainwright \cite{GW85} have shown 
that the expansion of the fluid limits to positive infinity as the IS is 
approached, so that the fluid must have non-zero expansion.  

We have now shown that if the conditions of Theorem \ref{thm:zar} are 
satisfied, then the conditions of Theorem \ref{thm:CW83} are also satisfied.
  This means that we have only three possible models to consider.  Both the 
spherically symmetric Wyman models and the plane symmetric Collins-Wainwright 
models need to be eliminated from consideration, since neither has a geodesic 
fluid flow.  We will do this by proving below that these models do not, in fact,
 actually admit an IS, and thereby do not satisfy the conditions of Theorem \ref{thm:zar}.  This then leaves only the FRW models with their geodesic fluid flows and we are done.


\subsection{Spherically symmetric Wyman models}

In local comoving coordinates $(t,r,\theta,\phi)$, the metric of the 
spherically symmetric Wyman models \cite{CW83,Wyman46} has the form
\be
ds^2 = \frac{1}{U^2}\left[ -\ \frac{U^{'2}}{At\plus B}dt^2+dr^2
	+r^2(d\theta^2+\sin^2\theta d\phi^2) \right],
\end{equation}
where $U=U(v)\neq 0, v=t+r^2$, a prime denotes differentiation w.r.t.\ $v$, and 
\be
U''=U^2 \Leftrightarrow U^{'2}=\smallfrac{2}{3}U^3-\smallfrac{1}{4}A.
\end{equation}
The quantities $A$ and $B$ are constants satisfying $A^2+B^2\neq 0$, i.e.,\ it 
is not permitted that both $A=0$ and $B=0$.  The energy density, $\mu$, and the 
pressure, $p$, of the perfect fluid are given by,
\be
\mu = 3(Av\plus B)+12UU',\quad\quad\mu+p=\frac{20U^4}{3U'}.
\end{equation}

Let $\vec{u}$ be the unit flow vector of the fluid, so that 
$g(\vec{u},\vec{u})=-1$.  Since we are using comoving coordinates, $\vec{u}$ 
has the form
\be
\vec{u} = u^0 \frac{\partial}{\partial t}, \quad u^0>0.
\end{equation}
Thus $g_{00}u^0u^0=-1$.  Substituting for $g_{00}$ we obtain
\bea
-\ \frac{U^{'2}}{U^2}\ \frac{1}{At+B}(u^0)^2 &=& -1 \\
\Leftrightarrow\quad (u^0)^2 &=&\frac{U^2}{U^{'2}}\ (At+B).
\eea

If a spherically symmetric Wyman model has an IS, then each flow-line given by 
$r,\theta,\phi$ constant will encounter the IS at a particular value of $t$ 
(where $t$ could be infinite).  That is, $t=t(r)$ is the ``equation'' for the
 IS in the physical space-time.

We note, firstly, that the metric will only be Lorentzian when $At+B>0$, which 
will henceforth be a requirement.  This implies that:
\bea
\text{if} \quad A =0, &\text{then }& B>0 \qquad\ \ t\in (-\infty,\infty),\\
\text{if} \quad A >0, &\text{then }& t>-\smallfrac{B}{A} 
		\qquad t\in (-\smallfrac{B}{A},\infty),\\
\text{if} \quad A <0, &\text{then }& t<-\smallfrac{B}{A} 
		\qquad t\in (-\infty,-\smallfrac{B}{A}).
\eea

The expansion scalar $\theta$ for the fluid is given by
\be
\theta=	\left\{ \begin{array}{c} 
	\!\! -3\sqrt{At\plus B}\quad\text{if}\quad\frac{U}{U'}>0  \\[2mm]
	\ \ 3\sqrt{At\plus B}\quad\text{if}\quad\frac{U}{U'}<0.
	\end{array}\right.
\end{equation}

It is known \cite{GW85} that as we approach an IS along a flow-line, 
$\theta\rightarrow +\infty$.  From the expression for $\theta$ given above it 
is readily seen, therefore, that metrics with $A=0$ do not have an IS.  Also, 
for metrics with $A>0$ or $A<0$, no flow-line can encounter an IS at 
$t_0=t(r_0)$, where $t_0\in \R$, since the expansion scalar is not infinite 
there.
This leaves only two possibilities:


\begin{enumerate}
\item if $A>0$, the IS occurs at $t=+\infty$ along every flow-line, or
\label{case1}
\item if $A<0$, the IS occurs at $t=-\infty$ along every flow-line.
\label{case2} 
\end{enumerate}	

Since $U(v)\neq0$, then along any given flow-line either $U>0$ or $U<0$.  Also 
$U''=U^2>0$, so that $U'$ is a strictly monotonically increasing function of 
$t$ along each flow-line. 

We firstly consider case (\ref{case1}) above.
\begin{ind}		
It is known \cite{SSGVR} that as we approach an IS along a flow-line, 
$\mu+p\rightarrow +\infty$.  For the spherically symmetric Wyman models,
\be
\mu+p=\frac{20}{3}\frac{U^4}{U'}.
\end{equation}
Now suppose that, as $t\rightarrow +\infty$:

%
\begin{enumerate}
%
%
\item[a.] $U'\rightarrow-\beta$ or $0$, where $\beta\in\R^+$.  

This implies that $\mu+p<0$ as $t\rightarrow +\infty$, so that an IS cannot 
occur there.
%
%
\item[b.] $U'\rightarrow\beta$, where $\beta\in\R^+$.

From the equation $U^{'2}=\smallfrac{2}{3}U^3-\smallfrac{1}{4}A$, it follows 
that 
\be
U\rightarrow 	\left[\frac{3}{2}(\beta^2\plus\smallfrac{1}{4}A) 
		\right]^\frac{1}{3}.
\end{equation}
Thus
\be
\mu+p\rightarrow \frac{20}{3}\ 
		\frac{	\left[\frac{3}{2}(\beta^2\plus\smallfrac{1}{4}A) 
			\right]^\frac{4}{3}}{\beta} \in\R 
		\quad \text{ as}\quad t\rightarrow +\infty,
\end{equation}
so that an IS cannot occur there. 
%
%
\item[c.] $U'\rightarrow +\infty$.

This implies that $U$ is a strictly monotonically increasing function of $t$ 
along the flow-line as $t\rightarrow +\infty$.  Indeed, from the constraint equation 
$U^{'2}=\smallfrac{2}{3}U^3-\smallfrac{1}{4}A$, it may be seen that 
$U\rightarrow +\infty$ as $t\rightarrow +\infty$.  This case requires further 
examination.  Using the asymptotic relationship
\be
U' \approx \frac{\sqrt{2}}{\sqrt{3}}\ U^{\frac{3}{2}},
\end{equation}
we obtain the following asymptotic relationships:
\bea
\mu &\approx& 3At+12\smallfrac{\sqrt{2}}{\sqrt{3}}\ U^{\frac{5}{2}} \\
\text{and  } \mu+p &\approx& 10\smallfrac{\sqrt{2}}{\sqrt{3}}\ U^{\frac{5}{2}}.
\eea

It is known \cite{SSGVR} that as we approach an IS with 
$\lambda=-\infty, \mu=o(\mu+p)$.  Since, in this case, 
$\mu>\mu+p$ as $t\rightarrow +\infty$, an IS with  
$\lambda=-\infty$ cannot occur there.

It is also known \cite{SSGVR} that as we approach an IS with 
$-\infty<\lambda<1$, $\mu+p\approx\smallfrac{2}{3}(2-\lambda)\mu$.  
In this case, the following asymptotic relationship would have to hold:
\bea
10\frac{\sqrt{2}}{\sqrt{3}}\ U^{\frac{5}{2}} &\approx& 
	2(2-\lambda)At+8\frac{\sqrt{2}}{\sqrt{3}}(2-\lambda)U^{\frac{5}{2}} \\
\Leftrightarrow \quad (2-\lambda)At &\approx&  
	\frac{\sqrt{2}}{\sqrt{3}}(4\lambda-3)U^{\frac{5}{2}}
\eea
Now $2-\lambda>1$ and $A>0$.  If $\lambda<\smallfrac{3}{4}$, then 
$4\lambda-3<0$, and it may be seen that the above asymptotic relationship 
cannot hold.  Thus an IS with $-\infty<\lambda<\smallfrac{3}{4}$ cannot 
occur as $t\rightarrow +\infty$.

If $\smallfrac{3}{4}<\lambda<1$, then
\bea
t &\approx& \alpha_\lambda U^{\frac{5}{2}}, 
		\quad\text{where}\quad\alpha_\lambda=
			\frac{\sqrt{2}}{\sqrt{3}}\ \frac{1}{A}\ 
			\frac{4\lambda\minus 3}{2\minus\lambda} \in\R^+ \\
\Leftrightarrow\quad U &\approx& 
		\left( \frac{1}{ \alpha_\lambda} \right)^\frac{2}{5}
			t^\frac{2}{5}.
\label{eq:aymprel}
\eea

Now $U'\approx\frac{\sqrt{2}}{\sqrt{3}}\ U^{\frac{3}{2}}$, so that  
\be
U'\approx\beta_\lambda t^{\frac{3}{5}}, \quad\text{ where }
\ \beta_\lambda=\frac{\sqrt{2}}{\sqrt{3}}
		\left(\frac{1}{\alpha_\lambda}\right)^\frac{3}{5}\in\R^+.
\end{equation}
Using L'H\^{o}pital's rule yields 
\be
U\approx\smallfrac{5}{8}\beta_\lambda t^{\frac{8}{5}}, 
\end{equation}
which is not consistent with the asymptotic relationship 
(\ref{eq:aymprel}) given above.   Thus 
an IS with $\smallfrac{3}{4}<\lambda<1$ cannot occur as $t\rightarrow +\infty$.

The only remaining possibility is that as $t\rightarrow +\infty$, we approach 
an IS with $\lambda=\smallfrac{3}{4}$.  For this case, $t=o(U^\frac{5}{2})$.  
It is known \cite{GW85} that as we approach an IS along a flow-line,

\be
\frac{C^{abcd}C_{abcd}}{R^{pq}R_{pq}}\rightarrow 0.
\end{equation}
In this case, 
\be
C^{abcd}C_{abcd}=\smallfrac{256}{3}r^4U^6,  
\end{equation}

\small{\bea
R^{pq}R_{pq} &=&\frac{\left(256U^6U^{'2}-768U^3U^{'4}
		+576U^{'6}\right)r^4}{U^{'2}} \nonumber\\
	&&\hugenegspace+ \left(192r^2U^3U^{'2}-288r^2U^{'4}
		-18AUU^{'}+216UU^{'3}\right. \nonumber\\
	&&\qquad\qquad\qquad\quad
		\left.-72U^4U^{'}+36(At+ B)U^{'2} \right) 
		\frac{(At+ B)}{U^{'2}}\nonumber \\
	&&\hugenegspace+\ \frac{\left(-\ 864UU^{'5}-48AU^4U^{'}
		+72AUU^{'3}-192 U^7U^{'}
			+ 864U^4U^{'3} \right)r^2}{U^{'2}} \nonumber\\
	&&\hugenegspace+\ \frac{48U^8-192U^5U^{'2}-48AU^2U^{'2}
		+336U^2U^{'4}+24AU^5+3A^2U^2}{U^{'2}}\ .\nonumber
\eea}
\normalsize
\be
\ee
One can analyse the asymptotic behaviour of $R^{pq}R_{pq}$ using the 
relationship
\be
U'=\smallfrac{\sqrt{2}}{\sqrt{3}}\ U^\frac{3}{2}\left[ 1-\frac{3}{8}
		\frac{A}{U^3}\right]^\frac{1}{2}.
\end{equation}
It is found that the largest terms, namely those of order $U^6$, cancel one 
another, as do the next largest terms, namely those of order $U^\frac{11}{2}$.
  Therefore, $R^{pq}R_{pq}=o(U^\frac{11}{2})$.  It follows that
\be
U^\frac{1}{2}=o\left( \frac{C^{abcd}C_{abcd}}{R^{pq}R_{pq}}  \right).
\end{equation}
Thus the condition that $\frac{C^{abcd}C_{abcd}}{R^{pq}R_{pq}}\rightarrow 0$ 
as $t\rightarrow +\infty$ does not hold, so that an IS with 
$\lambda=\smallfrac{3}{4}$ cannot occur there.
\end{enumerate}				
\end{ind}				

We conclude that for spherically symmetric Wyman models with $A>0$, no IS 
occurs at $t=+\infty$.


Now consider case (\ref{case2}), where $A<0$.  We will examine the 
possibility that an IS occurs at $t=-\infty$ along every flow-line.
\begin{ind}
Recall that as we approach an IS along a flow-line, 
$\mu+p\rightarrow +\infty$.  For the spherically symmetric Wyman models,
\be
\mu+p=\frac{20}{3}\frac{U^4}{U'}.
\end{equation}
Now suppose that, as $t\rightarrow -\infty$:
%
%
\begin{enumerate}

\item[a.] $U'\rightarrow -\infty$ or $-\beta$, where $\beta \in\R^+$.  This implies
 that $\mu+p<0$ as $t\rightarrow -\infty$, so that an IS cannot occur there.

\item[b.] $U'\rightarrow \beta$, where $\beta \in\R^+$.  From the equation 
$U^{'2}=\smallfrac{2}{3}U^3-\smallfrac{1}{4}A$, it follows that
\be
U\rightarrow \left[ \smallfrac{3}{2}(\beta^2+\smallfrac{1}{4}A) 
		\right]^\frac{1}{3}.
\end{equation}
Thus 
\be
\mu+p \rightarrow \frac{20}{3}\ \frac{ \left[ 
			\smallfrac{3}{2}(\beta^2+\smallfrac{1}{4}A) 
			\right]^\frac{4}{3}}{\beta} 
		\in \R\quad\text{as }\ t\rightarrow -\infty,
\end{equation}
so that an IS cannot occur there.

\item[c.] $U'\rightarrow 0^+$.  This implies that $U$ is a strictly 
monotonically increasing function of $t$ along the flow-line.  So we can set 
$U=\smallfrac{1}{2}(3A)^\frac{1}{3}+H(t)$, where $H(t)\rightarrow 0^+$ and 
$H'=U'>0$ as $t\rightarrow -\infty$.  Substituting these expressions for $U$ 
and $U'$ into the equation $U^{'2}=\smallfrac{2}{3}U^3-\smallfrac{1}{4}A$, 
we obtain
\bea
(H')^2 	&=& \frac{2}{3}\left[\frac{1}{2}(3A)^\frac{1}{3}+H \right]^3
		-\frac{1}{4}A \\
	&=& \frac{2}{3}\left[\frac{3}{8}A+\frac{3}{4}(3A)^\frac{2}{3}H
		+\frac{3}{2}(3A)^\frac{1}{3}H^2+H^3 \right]-\frac{1}{4}A.
\eea
Since $H=o(1)$ as $t\rightarrow -\infty$, this implies that, as 
$t\rightarrow -\infty$,
\bea
(H')^2 	&\approx& \frac{1}{2}(3A)^\frac{2}{3}H\\
\Leftrightarrow\quad H' &\approx& 
			-\ \frac{1}{\sqrt{2}}(3A)^\frac{1}{3}H^\frac{1}{2}
				\quad\text{(since $A<0$)}\\
\Leftrightarrow\quad \frac{H'}{H^\frac{1}{2}} &\approx& 
			-\ \frac{1}{\sqrt{2}}(3A)^\frac{1}{3}.
\end{eqnarray}

Using L'H\^{o}pital's rule yields
\be
2H^\frac{1}{2}\approx -\ \frac{1}{\sqrt{2}}(3A)^\frac{1}{3}t,
\end{equation}

which cannot be true since $2H^\frac{1}{2}\rightarrow 0^+$ but 
$-\frac{1}{\sqrt{2}}(3A)^\frac{1}{3}t\rightarrow -\infty$ as  
$t\rightarrow -\infty$.  We conclude that 
the asymptotic behaviour $U'\rightarrow 0^+$ as $t\rightarrow -\infty$ is not 
consistent with the constraint equation 
$U^{'2}=\smallfrac{2}{3}U^3-\smallfrac{1}{4}A$, 
which must be satisfied by these models.

\end{enumerate} 
\end{ind}		

Thus for spherically symmetric Wyman models with $A<0$, no IS occurs at 
$t=-\infty$.

This then proves that the spherically symmetric Wyman models do not admit an 
isotropic singularity.


\subsection{Plane symmetric Collins-Wainwright models}

In local comoving coordinates $(t,x,y,z)$, the metric of the plane symmetric 
Collins-Wainwright models \cite{CW83} has the form
\be
ds^2 = \frac{C^2}{U^2}	\left[ 
			-\ \frac{U^{'2}}{m^2}dt^2+dx^2+e^{-2x}(dy^2+dz^2)
			\right],
\end{equation}
where $U=U(v)\neq 0, v=t+x$, a prime denotes differentiation w.r.t.\ $v$, and 
\be
U''+U'=-U^2.
\end{equation}
The quantities $C$ and $m$ are positive constants.  The energy density,
 $\mu$, and the pressure, $p$, of the perfect fluid are given by,
\be
\mu = \frac{1}{C^2}\left[3m^2-2U^3-3(U'\plus U)^2 \right],
		\quad\quad\mu+p=\frac{2U^4}{C^2U'}.
\end{equation}

Let $\vec{u}$ be the unit flow vector of the fluid, so that 
$g(\vec{u},\vec{u})=-1$.  Since we are using comoving coordinates, 
$\vec{u}$ has the form
\be
\vec{u} = u^0 \frac{\partial}{\partial t}, \quad u^0>0.
\end{equation}
Thus $g_{00}u^0u^0=-1$.  Substituting for $g_{00}$ we obtain
\bea
-\ \frac{C^2}{m^2}\ \frac{U^{'2}}{U^2}(u^0)^2 &=& -1 \\
\Leftrightarrow\quad (u^0)^2 &=&\frac{m^2}{C^2}\frac{U^2}{U^{'2}}.
\eea

The expansion scalar $\theta$ for the fluid is given by
\be
\theta=	\left\{\begin{array}{c}
	\ 3\frac{m}{C}	\quad\text{if}\quad\frac{U}{U'}>0 \\ [2mm]
	- 3\frac{m}{C}	\quad\text{if}\quad\frac{U}{U'}<0.
	\end{array}\right.
\end{equation}
The flow-lines of the fluid are given by $x=$ constant, $y=$ constant, and 
$z=$ constant.  It is known \cite{GW85} that 
as we approach an IS along a flow-line, $\theta\rightarrow+\infty$.  Since, for
 the plane symmetric Collins-Wainwright models the expansion scalar is either
 a positive constant or a negative constant along each flow-line, it is clear 
that the condition $\theta\rightarrow+\infty$ is never satisfied along a 
flow-line.  Thus the plane symmetric Collins-Wainwright models do not admit an
isotropic singularity.

\end{proof}	


\section{General proof of the Zero Acceleration Result}
\label{sec:general_proof_of_zar}
The proof by elimination of the ZAR utilises the theorem of Collins and 
Wainwright \cite{CW83}.  Given the existence of the proof by elimination, 
there should also
exist a general proof, which we now give.  We believe that the
general proof may also provide direction on how to proceed in proving the FRW 
conjecture (see the discussion in Section \ref{sec:zar_discussion}).
\pagebreak

\begin{proof}{General proof of the ZAR}

In order to prove Theorem \ref{thm:zar}, we will show that, for a space-time
 which satisfies the conditions of the theorem, the electric and 
magnetic parts of the Weyl tensor must be zero.  A barotropic perfect fluid 
cosmology with zero Weyl tensor must be a FRW model, and hence the fluid flow
 is geodesic.

The General Vorticity Result of Scott \cite{SSGVR} ensures that a space-time 
which satisfies the conditions of Theorem \ref{thm:zar} must be 
irrotational --- i.e., the vorticity, $\omega^2$, of the fluid must satisfy
$\omega^2=0$.  Furthermore, a space-time satisfying the conditions of 
Theorem \ref{thm:zar} is also shear-free, i.e., $\sigma^2=0$.

The vorticity, $\omega^2$, and the shear, $\sigma^2$, of the fluid are related 
to the vorticity, $\s\omega^2$, and the shear, $\s\sigma^2$, of the unphysical 
fluid flow, $\s \vec{u}$, in the unphysical space-time $(\s\M,\s g)$ by
\begin{equation}
\begin{array}{ll}
\qquad\:\s \omega^2 = \Omega^2 \omega^2,\quad &\qquad\: \s\sigma^2 = \Omega^2\sigma^2 \\
\implies \quad \s \omega^2  	=0	&\implies\quad\s\sigma^2=0 
\end{array}
\end{equation}

The constraint equation \cite[p130 eqn 4.19]{Ellis71} for the magnetic part 
of the unphysical Weyl tensor is given by
\be
\s H_{ad}= 2\s\dot{u}_{(a}\s\omega_{d)}-\s h_{at}\s h_{ds}\left(\cds{\s{\omega^{(t}}_b}{c} +\cds{\s{\sigma^{(t}}_b}{c} \right) \s\eta^{s)fbc}\s u_f ,
\end{equation}
which reduces to 
\be
\s H_{ad}= 0.
\end{equation}
The physical and unphysical magnetic parts of the Weyl tensor are related by
\be
H_{ab}=\s H_{ab},
\end{equation} 
and thus the magnetic part of the Weyl tensor for the physical space-time is
 zero as required.

The Einstein tensor for $(\M,g)$ can be decomposed relative to the fluid 
flow, $\vec{u}$, as follows:
\be
G_{ab}=Au_au_b+Bh_{ab}+\Sigma_au_b+\Sigma_bu_a+\Sigma_{ab},
\end{equation}
where
\be
\begin{array}{ll}
\: A = G_{cd}u^cu^d,	&\qquad\:\ B = \smallfrac{1}{3}h^{cd}G_{cd}, \\
\Sigma_a = -h_{ac}R^{cd}u_d,	
		&\qquad \Sigma_{ab} = h_{ac}R^{cd}h_{bd}
					-\smallfrac{1}{3}h_{ab}R^{cd}h_{cd},
\end{array}
\end{equation}
and
\be
h_{ab}= g_{ab}+u_au_b.
\end{equation}

A perfect fluid cosmology satisfying the conditions of Theorem \ref{thm:zar} is irrotational, and the unit timelike fluid congruence is regular at an isotropic singularity.  We have the freedom to choose a cosmic time function $T$ defined on $\s \M$, and a conformal factor $\Omega(T)$, such that the unphysical fluid flow $\s \vec{u}$ in $\s \M$ is orthogonal to the spacelike hypersurfaces $T=$ constant.  Using such a cosmic time function, we will employ comoving normal coordinates $(T,x^\alpha)$ based on the hypersurface $T=0$ in $\s \M$.  The hypersurface $T=0$ in $\s \M$ is referred to as the isotropic singularity.

A perfect fluid satisfies the field equation $\Sigma_{ab}=0$.  Thus, in normal coordinates (\cite[p108]{GW85}, also equation (\ref{eq:Sigma_sigma})), the unphysical counterpart $\s\Sigma_{ab}$ is given by the equation
\bea
\s\Sigma_{ab}		&=& -\ 2F\frac{\Omega'}{\Omega}\s\sigma_{ab} \\ 
\implies\quad \s\Sigma_{ab}	&=& 0.
\eea

Written in terms of geometrical quantities, the unphysical shear propagation equation \cite[p29]{Ellis73} is given by
\bea
\lefteqn{\s{h_r}^a\s{h_s}^c\left( \s\dot{\sigma}_{ac}-\cds{\s\dot{u}_{(a}}{c)} \right) + \smallfrac{1}{3}\s h_{rs}\left(\cds{\s\dot{u}^a}{a} -2\s\sigma^2-\s\omega^2 \right)  -\smallfrac{1}{2}\s\Sigma_{rs}} \qquad\qquad\quad\nonumber\\ 
& &-\s\dot{u}_r\s\dot{u}_s +\s\sigma_{rd}\s{\sigma^d}_s +\smallfrac{2}{3}\s\theta\s\sigma_{rs} +\s\omega_r\s\omega_s +\s E_{rs}=0,
\eea
which, in the present situation, reduces to 
\be
-\s{h_r}^a\s{h_s}^c\cds{\s\dot{u}_{(a}}{c)} + \smallfrac{1}{3}\s h_{rs}\cds{\s\dot{u}^a}{a}-\s\dot{u}_r\s\dot{u}_s+\s E_{rs}=0 .
\end{equation}
The $(0a)$ equations are trivial, so setting $r=\mu, s=\nu$ and raising these indices, and noting that in normal coordinates $\s h^{a0}=0$, we obtain the equation
\be
-\s h^{\mu\alpha}\s h^{\nu\beta}\cds{\s\dot{u}_{(\alpha}}{\beta)} + \smallfrac{1}{3}\s h^{\mu\nu}\cds{\s\dot{u}^a}{a}-\s\dot{u}^\mu\s\dot{u}^\nu+\s E^{\mu\nu}=0 .
\label{eq:specialised_shear_prop}
\end{equation}

The Bianchi identity dealing with the propagation of the electric part of the unphysical Weyl tensor is given \cite[p131 eqn 4.21d using the Appendix p179 eqn 4.21d]{Ellis71} by the equation
\bea
&&\hugenegspace \s{h_a}^m\s{h_c}^t\s\dot{E}^{ac}+\s{h_a}^{(m}\s\eta^{t)rsd}\s u_r\cds{{\s H^a}_s}{d}-2\s{H_q}^{(t}\s\eta^{m)bpq}\s u_b\s\dot{u}_p\nonumber\\
&&\hugenegspace\ +\s h^{mt}\s\sigma^{ab}\s E_{ab}+\s\theta\s E^{mt}-3\s{E_s}^{(m}\s\sigma^{t)s}-\s{E_s}^{(m}\s\omega^{t)s}= \nonumber\\
&&\hugenegspace\medspace-\smallfrac{1}{2}(\s A \plus \s B)\s\sigma^{tm}-\s\dot{u}^{(t}\s\Sigma^{m)}-\smallfrac{1}{2}\s h^{ta}\s h^{mc}\cds{\s\Sigma_{(a}}{c)}-\smallfrac{1}{2}\s{h^t}_a\s{h^m}_c\s \dot{\Sigma}^{ac}\nonumber\\
&&\hugenegspace\medspace-\smallfrac{1}{2}\s\Sigma^{b(m}\s{\sigma_b}^{t)}-\smallfrac{1}{2}\s\Sigma^{b(m}\s{\omega_b}^{t)}-\smallfrac{1}{6}\s\Sigma^{tm}\s\theta+\smallfrac{1}{6}\left(  \cds{\s\Sigma^a}{a} \plus \s\dot{u}_a\s\Sigma^a \plus \s\Sigma^{ab}\s\sigma_{ab} \right)\s h^{mt},
\eea
which, in the present situation, reduces to 
\be
\s{h_a}^m\s{h_c}^t\s \dot{E}^{ac}+\s\theta\s E^{mt}= -\s\dot{u}^{(t}\s\Sigma^{m)}-\smallfrac{1}{2}\s h^{ta}\s h^{mc}\cds{\s\Sigma_{(a}}{c)}+\smallfrac{1}{6}\left(  \cds{\s\Sigma^a}{a} \plus \s\dot{u}_a\s\Sigma^a \right)\s h^{mt}.
\label{eqn:reduced_bianchi}
\end{equation}

Since a perfect fluid satisfies the field equation $\Sigma^a=0$, in normal coordinates (\cite[p108]{GW85}, also equation (\ref{eq:Sigma_u})), the unphysical counterpart $\s\Sigma^a$ is given by the equation
\be
\s\Sigma^a = 2F\frac{\Omega'}{\Omega}\s\dot{u}^a,
\end{equation}
and thus the terms on the right hand side of the reduced Bianchi identity (equation (\ref{eqn:reduced_bianchi})) become
\bea
\s\dot{u}^{(t}\s\Sigma^{m)} &=& 2F\frac{\Omega'}{\Omega}\s\dot{u}^t\s\dot{u}^m \\
\cds{\s\Sigma_{(\alpha}}{\beta)} &=& -\ 2F\frac{\Omega'}{\Omega}\s\dot{u}_\alpha\s\dot{u}_\beta+ 2F\frac{\Omega'}{\Omega}\cds{\s\dot{u}_{(\alpha}}{\beta)} \\
\cds{\s\Sigma^a}{a}  &=& -\ 2F\frac{\Omega'}{\Omega}\s\dot{u}^a\s\dot{u}_a + 2F\frac{\Omega'}{\Omega}\cds{\s\dot{u}^a}{a}  \\
\s\dot{u}_a\s\Sigma^a &=& 2F\frac{\Omega'}{\Omega}\s\dot{u}^a\s\dot{u}_a .  
\eea

For the reduced Bianchi identity, the $(0a)$ equations are trivial, so setting $m=\mu, t=\nu$, the Bianchi identity further simplifies to the equation
\be
\s{h_\alpha}^\mu\s{h_\beta}^\nu\s \dot{E}^{\alpha\beta}+\s\theta\s E^{\mu\nu}= F\frac{\Omega'}{\Omega}\left( -\s\dot{u}^\mu\s\dot{u}^\nu -\s h^{\alpha\nu}\s h^{\beta\mu}\cds{\s\dot{u}_{(\alpha}}{\beta)}+\smallfrac{1}{3}\cds{\s\dot{u}^a}{a}\s h^{\mu\nu}  \right) .
\end{equation}
Now inserting the shear propagation equation (\ref{eq:specialised_shear_prop}), and noting that in normal coordinates $\s {h_\alpha}^\mu={\delta_\alpha}^\mu$, we obtain the very simple equation
\bea
& &\s \dot{E}^{\mu\nu}+\s\theta\s E^{\mu\nu} = -F\frac{\Omega'}{\Omega}\s E^{\mu\nu} \\
\implies& & {\s E^{\mu\nu}}_{,0}+\s{\Gamma^\mu}_{0\alpha}\s E^{\alpha\nu}+\s{\Gamma^\nu}_{0\alpha}\s E^{\alpha\mu}+\left( \frac{\s\theta}{F}+\frac{\Omega'}{\Omega} \right)\s E^{\mu\nu} = 0 .
\label{eq:dEd0withgammas}
\eea

In normal coordinates,
\bea
\s{\Gamma^\mu}_{0\alpha} = \smallfrac{1}{2}\s g^{\mu\gamma}\s g_{\gamma\alpha,0}\medspace,\quad &\quad& \quad \s{\theta^\mu}_\alpha = \smallfrac{F}{2}\s g^{\mu\gamma}\s g_{\gamma\alpha,0}\\
\implies\quad  \s{\Gamma^\mu}_{0\alpha} &\!\!=\!\!& \frac{\s{\theta^\mu}_\alpha}{F}.
\eea
We also know that
\bea
\s{\theta^\mu}_\alpha 	&=&\s{\sigma^\mu}_\alpha+\smallfrac{1}{3}\s\theta\s{h^\mu}_\alpha \\
			&=&\smallfrac{1}{3}\s\theta\s{h^\mu}_\alpha. 
\eea
Thus
\be
\s{\Gamma^\mu}_{0\alpha}=\frac{\s\theta}{3F}{\delta^\mu}_\alpha.
\end{equation}
Inserting this expression for $\s{\Gamma^\mu}_{0\alpha}$ into equation (\ref{eq:dEd0withgammas}) yields the equation
\be
{\s E^{\mu\nu}}_{,0}+\left( \frac{5\s\theta}{3F}+\frac{\Omega'}{\Omega} \right)\s E^{\mu\nu} = 0,
\end{equation}
from which we readily obtain the equation
\be
(\Omega \s E^{\mu\nu})_{,0}+ \frac{5\s\theta}{3F}(\Omega\s E^{\mu\nu}) = 0.
\end{equation}

In order to solve this system of partial differential equations, we first need to obtain some initial conditions.  From Goode and Wainwright \cite[p108]{GW85} we know that
\bea
\s{\Sigma_a}^b|_{T=0}	&=&\frac{2}{3\minus\lambda}\s{{}^3\negthinspace S_a}^b|_{T=0} \quad\text{and}\\
\nonumber \\
\s{E_a}^b|_{T=0}	&=&\frac{2\minus\lambda}{3\minus\lambda}\s{{}^3\negthinspace S_a}^b|_{T=0} \ ,
\eea
where $\s{{}^3\negthinspace S_a}^b|_{T=0}$ is the trace-free Ricci tensor of the isotropic singularity (i.e., the hypersurface $T=0$ in $\s\M$).  Since $\s{\Sigma_a}^b=0$, it follows that
\be
\s{{}^3\negthinspace S_a}^b|_{T=0} =0,
\end{equation}
and thus
\bea
\s{E_a}^b|_{T=0} &=&0 \\
\implies\quad \s E^{ab}|_{T=0} &=&0 \\
\implies\quad (\Omega\s E^{\mu\nu})|_{T=0} &=&0\ . \label{eq:zar_init_conditions}
\eea

Since we are using comoving coordinates, the coordinates $x^\alpha$ are 
constant along a 
flow-line, and thus we can treat the partial differential equations as 
ordinary differential equations, in the variable $T$, along the flow-line.  
If $\Omega \s E^{\mu\nu} $ was defined on an 
open neighbourhood of $T=0$, we could then use a standard existence/uniqueness 
theorem (see Appendix \ref{appendix:ODE}) to show that 
$\Omega \s E^{\mu\nu} \equiv0$.  However, 
$\Omega \s E^{\mu\nu} $ is only defined on the interval $[0,b)$, where $b$ is some positive constant.  Suppose 
then that $\Omega \s E^{\mu\nu} \neq 0$ anywhere on $(0,b)$.  (If $\Omega \s E^{\mu\nu} =0$ at any point in $(0,b)$, then 
we could apply the standard existence/uniqueness theorem to obtain 
$\Omega \s E^{\mu\nu} \equiv 0$ on $[0,b)$).  Let $a\in(0,b)$ and set 
$(\Omega \s E^{\mu\nu})(a)=(\Omega \s E^{\mu\nu})_a \neq 0$.
Then
\bea
\frac{(\Omega \s E^{\mu\nu})_{,0}}{(\Omega \s E^{\mu\nu})}	
	&=& -\ \frac{5\s\theta}{3F} 
		\quad(\Omega \s E^{\mu\nu} \neq 0\text{ by assumption})\\
\implies\quad (\Omega \s E^{\mu\nu})(T)	&=& (\Omega \s E^{\mu\nu})_a\ e^{-\int_a^T\frac{5\s\theta}{3F}d\tau }.	
\eea
Now since $F$ and $\s\theta$ are, respectively, at least $C^3$ and $C^2$ and $F\neq 0$ on an open neighbourhood of $T=0$,
\be
\int_a^T\frac{5\s\theta}{3F}\ d\tau=O(1)\quad\text{as }\ T\rightarrow0^+.
\end{equation}
It follows that $e^{-\int_a^T\frac{5\s\theta}{3F}d\tau}\nrightarrow 0$ as $T\rightarrow0^+$, and since $(\Omega \s E^{\mu\nu})_a \neq 0$,
\be
(\Omega \s E^{\mu\nu}) \nrightarrow 0\quad\text{as }\ T\rightarrow0^+,
\end{equation}
which contradicts the above initial condition (equation (\ref{eq:zar_init_conditions})).  Hence, along the flow-line,
\bea
\Omega \s E^{\mu\nu} 		&=&0\\
\implies\quad \s E^{\mu\nu} 	&=&0\quad\text{since }\ \Omega>0 \text{ on } (0,b)\\
\implies\quad \s E_{ab} 	&=&0\ .
\eea 

The physical and unphysical electric parts of the Weyl tensor are related by
\bea
E_{ab}&=&\s E_{ab} \\
\implies\quad E_{ab}&=& 0.
\eea
Since $E_{ab}=0$ along every flow-line, the electric part of the Weyl tensor for the physical space-time is zero, as required.

\end{proof}				

 
\section{Discussion}
\label{sec:zar_discussion}
It is interesting to note that there may exist barotropic perfect fluid cosmological models, with unit timelike fluid congruence which is regular at an IS, which do \emph{not} satisfy the dominant energy condition, yet which are irrotational.  Such models can be accommodated in an alternative version of Theorem \ref{thm:zar}, given as Theorem \ref{thm:zar_alternative}.  Neither the proof of the ZAR by elimination (Section \ref{sec:proof_of_zar_elimination}), nor the general proof of the ZAR (Section \ref{sec:general_proof_of_zar}) use the dominant energy condition, except, in so far as it is needed to prove the GVR of Scott \cite{SSGVR}.  We can therefore remove the dominant energy condition assumption from the statement of Theorem \ref{thm:zar} (i.e., $-1<\lambda<1$) and instead replace it with the assumption that the fluid flow is irrotational, and the two proofs will proceed exactly as before.

\begin{thm}[Alternative Zero Acceleration Result]
\label{thm:zar_alternative}
If a space-time $(\M,g)$:
\begin{itemize}
\item is a $C^3$ solution of the Einstein field equations (EFE), 
\item has barotropic perfect fluid source, 
\end{itemize}
and the unit timelike fluid congruence \vec{u} is:
\begin{itemize}
\item irrotational,
\item shear-free,
\item regular at an isotropic singularity, 
\end{itemize}
then the fluid flow is necessarily geodesic.
\end{thm}

\begin{figure}[!t]
\begin{center}
\includegraphics[scale=.4]{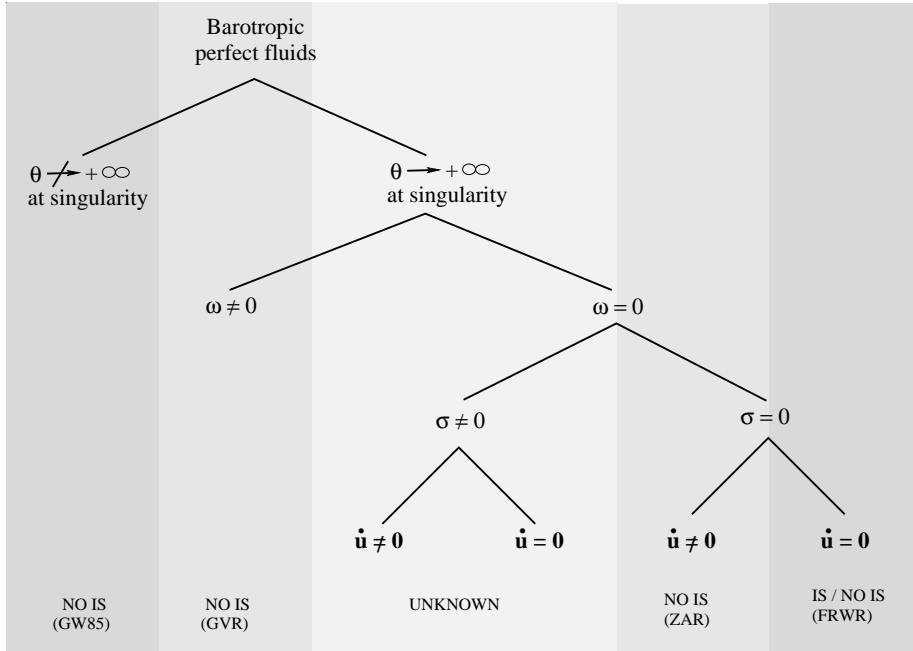}
\end{center}
\caption{Known results about barotropic perfect fluids}
\label{fig:tree_diagram}
\end{figure}

Perfect fluid space-times can be categorised according to their vorticity, shear, acceleration, and expansion.  Using this categorisation, all currently known results about barotropic perfect fluid cosmologies with a unit timelike fluid congruence which is regular at an isotropic singularity are summarised in Figure \ref{fig:tree_diagram}. 

The abbreviations used in Figure \ref{fig:tree_diagram} are:
\begin{description}
\itemsep=0pt
\parsep=0pt
\item[GW85]Goode and Wainwright's 1985 paper \cite{GW85} on Isotropic Singularities
\item[GVR]The General Vorticity Result \cite{SSGVR} due to Scott (note: $-1<\lambda<1$)
\item[ZAR]The Zero Acceleration Result 
\item[FRWR]The FRW Result \cite{FRWR}
\item[NO IS] The unit timelike fluid congruence is not regular at an isotropic singularity
\end{description}

Figure \ref{fig:tree_diagram} shows that much is already known about barotropic perfect fluid cosmologies in relation to isotropic singularities.  Goode and Wainwright \cite{GW85} have shown that the expansion of the fluid must approach infinity at an isotropic singularity, and hence, whenever the expansion of the fluid does not approach infinity at the initial singularity, there can be no IS.  Scott \cite{SSGVR} has shown that for barotropic perfect fluid cosmologies satisfying the dominant energy condition, if the space-time has an IS, then the vorticity of the fluid is zero.  Hence, whenever the vorticity is non-zero, there can be no IS.  We have just shown (ZAR) that if the fluid is irrotational, shear-free, and not geodesic, then there can be no IS.  In a forthcoming paper \cite{FRWR} we find the precise necessary and sufficient conditions for a Friedmann-Robertson-Walker model to admit an IS.  The only remaining unknown cases, therefore, are for irrotational perfect fluids with non-zero shear.

For the case of a barotropic, irrotational, perfect fluid cosmology with non-zero shear, there are examples of cosmologies with, and without, geodesic fluid flow, both with, and without, isotropic singularities.  For example, the Kantowski-Sachs models 
\cite{KS66}
are irrotational, have non-zero shear and geodesic fluid flow and have an IS \cite{GW85}.  The cosmologies given by Mars \cite{Mars95} are irrotational, have non-zero shear, and are not geodesic, some of which have an IS.  A review of known examples of cosmologies which admit an IS is given in \cite{potsdam}.

An idea which has often been linked with the concept of an isotropic singularity \cite{GW85,Tod87,Scott89,GCW92,Newman93} is the Weyl curvature hypothesis \cite{Penrose79}.  Non-rigorously stated, the Weyl curvature hypothesis says that ``the natural thermodynamic boundary condition for the universe is that the Weyl tensor should vanish at any initial singularity''.  The examples given in the previous paragraph do not satisfy the Weyl curvature hypothesis which leads us to what is known as the FRW conjecture \cite{Tod87,Scott89,GW85,GCW92}.

\pagebreak
\begin{conj}[FRW]
If a space-time $(\M,g)$ is:
\begin{enumerate}
\item a $C^3$ solution of the Einstein field equations with a barotropic 
	perfect fluid source, and
\item the unit timelike fluid congruence \vec{u} is regular at an isotropic 
	singularity (with $-1<\lambda<1$), and
\item the Weyl curvature hypothesis holds,
\end{enumerate}
then the space-time is necessarily a Friedmann-Robertson-Walker model.
\end{conj}


\section*{Acknowledgements}
The tensors in this paper were calculated using Maple with GRTensorII.


\appendix
\section*{Appendices}			

\section{Definition of an Isotropic Singularity}
\label{ap:defns}
In 1985 Goode and Wainwright \cite{GW85} introduced the concept of the 
isotropic singularity (IS).  Scott \cite{Scott88,SSGVR} has amended their 
original definition to remove some inherent redundancy, and this amended 
definition of an isotropic singularity is given in Definition \ref{def:IS}. 
\begin{defn}[Isotropic singularity]
\label{def:IS}
A space-time $(\M,g)$ is said to admit an \emph{isotropic singularity} if 
there exists a space-time $(\s\M, \s g)$, a smooth cosmic time function $T$ 
defined 
on $\s\M$, and a conformal factor $\Omega (T)$ which satisfy
\begin{enumerate}
\item $\M$ is the open submanifold $T>0$,
\item $g= \Omega^{2}(T) \s g$  on $\M$, with $\s g$ regular (at least $C^{3}$ 
	and non-degenerate) on an open neighbourhood of $T=0$,
\item $\Omega(0) = 0$ and $\exists\ b>0$ such that $\Omega \in C^{0}[0,b] \cap  
	C^{3}(0,b]$ and $\Omega(0,b] >0$,
\item $\lambda \equiv \lim_{T \rightarrow 0^{+}} L(T)$ exists, $\lambda\neq 1$, 
	where $L \equiv \frac{\Omega''}{\Omega}{\left( \frac{\Omega}{\Omega'} 
	\right)}^{2}$ and a prime denotes differentiation with respect to T.
\end{enumerate}
\end{defn} 
\begin{defn}[Unphysical flow]
With any unit timelike congruence $\vec{u}$ in $\M$ we can associate a unit 
timelike congruence $\s \vec{u}$ in $\s \M$ such that
\be
\s u^a = \Omega u^a \quad \text{in}\:\ \M.
\end{equation}
(a) If we can choose $\s \vec{u}$ to be regular (at least $C^3$) on an open 
	neighbourhood of $T=0$ in $\s \M$, we say that $\vec{u}$ is 
	\emph{regular at the isotropic singularity}, and \\
(b) if, in addition, $\s \vec{u}$ is orthogonal to $T=0$, we say that $\vec{u}$ 
	is \emph{orthogonal to the isotropic singularity}.
\end{defn}

\section{Existence/Uniqueness Theorem}
\label{appendix:ODE}
The following existence/uniqueness theorem can be found in \cite{ODE}.
\begin{thm}[Existence/uniqueness]
If the functions $p$ and $g$ are continuous on an open interval $\alpha<x<\beta$
 containing the point $x=x_0$, then there exists a unique function $y=\phi(x)$ 
that satisfies the differential equation
\be
y'+p(x)y=g(x)
\end{equation}
for $\alpha<x<\beta$, and that also satisfies the initial condition
\be
y(x_0)=y_0
\end{equation}
where $y_0$ is an arbitrary prescribed initial value.
\end{thm}

\section{Comoving Normal Coordinates}
We set up comoving normal coordinates $(T,x^\alpha)$ in $(\s\M,\s g)$, based on the hypersurface $T=0$ (see Section 3).  Thus
\be
\s g^{0\alpha} = 0,
\end{equation}
which implies that
\be
\s {\Gamma^\mu}_{0\alpha} = \frac{1}{2} {\s g}^{\mu \nu} {\s g}_{\nu \alpha,0}.
\end{equation}
Also
\be
\s u_a = -F^{-1}T_{,a}.
\end{equation}
Then,
\be
\begin{array}{lll}
{\s h}_{00} = 0  & {\s h}^{00} = 0  & {{\s h}_{0}}^{0} = 0 \\
{\s h}_{0 \alpha} = {\s h}_{\alpha 0} = 0 & {\s h}^{0 \alpha} = {\s h}^{\alpha 0} = 0 \quad & {{\s h}^{0}}_{\alpha} = {{\s h}^{\alpha}}_{0} = 0 \\
{\s h}_{\alpha \beta} = {\s g}_{\alpha \beta}  & {\s h}^{\alpha \beta} = {\s g}^{\alpha \beta}  & {{\s h}_{\alpha}}^{\beta} = {\delta_{\alpha}}^{\beta} 
\end{array}
\ee
\be
\begin{array}{ll}
\s {\dot{u}_{0}} = 0 &\quad \s {\dot{u}^{0}} = 0 \\
\s {\dot{u}_{\alpha}} = - \frac{F_{,\alpha}}{F} &\quad \s {\dot{u}^{\alpha}} = -\s g^{\alpha \beta} \frac{F_{,\beta}}{F}  
\end{array}
\ee
\bea
\s E_{a0} &=& 0 \\
{\s \theta}_{a0} &=& 0 \\
{\s \theta}_{\alpha \beta} &=& \frac{F}{2}{\s g}_{\alpha \beta ,0} 
\eea

Since the physical space-time has a perfect fluid source then \cite{GW85},
\bea
\label{eq:Sigma_sigma} \s \Sigma_{ab}=-2F\frac{\Omega'}{\Omega}\s\sigma_{ab},\\
\label{eq:Sigma_u} \s \Sigma^a=2F\frac{\Omega'}{\Omega}\s\dot{u}^a.
\eea



\newcommand{\journal}[7]{#1  (#5)  \textit{#3}, \textbf{#4}, #6-#7.}

\newcommand{\proc}[8]{#1 (#6) ``#2''  \textit{#3}  ed. #4 (#5) pp#7-#8.}

\newcommand{\book}[4]{#1 (#4) \textit{``#2''} (#3).}

\newcommand{\phd}[3]{#1 (#3) \textit{Ph.D. thesis} #2.}

\newcommand{\AOP}{A.P.}
\newcommand{\CQG}{Class.\ Quantum Grav.}
\newcommand{\GRG}{Gen.\ Rel.\ Grav.}
\newcommand{\JMP}{J.\ Math.\ Phys.}
\newcommand{\PRD}{Phys.\ Rev.\ D}
\newcommand{\PR}{Phys.\ Rev.}
\newcommand{\PRSLA}{Proc.\ R.\ Soc.\ Lond.\ A}



\end{document}